# Electron doping of the layered nickelate La$_4$Ni$_3$O$_{10}$ by aluminum substitution: A combined experimental and DFT study

Manimuthu Periyasamy[a], Lokanath Patra[b], Øystein S. Fjellvåg[c], Ponniah Ravindran[b], Magnus H. Sørby[c], Susmit Kumar[a], Anja O. Sjåstad[a] and Helmer Fjellvåg[a,*]

[a]Centre for Materials Science and Nanotechnology, Department of Chemistry, University of Oslo, PO Box 1033, N-0315 Oslo Norway

[b]Department of Physics, Central University of Tamil Nadu, Nilakkudi, Tamil Nadu 610005, India

[c]Department for Neutron Materials Characterization, Institute for Energy Technology, PO Box 40, NO-2027, Kjeller, Norway

*E-mail: *helmer.fjellvag@kjemi.uio.no*

The physical properties of La$_4$Ni$_3$O$_{10}$ with a 2D-like Ruddlesden-Popper-type crystal structure are extraordinarily dependent on temperature and chemical substitution. By introducing Al$^{3+}$ atoms randomly at the Ni-sites, the average oxidation state for the two non-equivalent Ni-cations is tuned and adopt values below the average of +2.67 in La$_4$Ni$_3$O$_{10}$. La$_4$Ni$_{3-x}$Al$_x$O$_{10}$ is a solid solution for $0.00 \leq x \leq 1.00$, and are prepared by the citric acid method, with attention paid on compositional control and homogeneity at low Al-level ($x$)-concentrations. The samples adopt a slightly distorted monoclinic structure ($P2_1/a$), evidenced by peak broadening of the (117) reflection. We report on a remarkable effect on the electronic properties induced by tiny amounts of homogeneously distributed Al-cations, with clear correspondence between resistivity, magnetization, diffraction and DFT data. DFT shows that electronically there is no significant difference between the non-equivalent Ni atoms and no tendency towards any Ni$^{3+}$/Ni$^{2+}$ charge ordering. The electron doping via Al-substitution has a profound effect on electric and magnetic properties. The resistivity changes from metallic to semiconducting with increasing band-gap at higher Al-levels, consistent with results from DFT. The metal-to-metal transition reported for La$_4$Ni$_3$O$_{10}$, which is often interpreted as a charge density wave, is maintained until $x = 0.15$ Al-level. However, the temperature characteristics of the resistivity change already at very low Al-levels ($x \leq 0.03$). A coupling of the metal-to-metal transition to the lattice is evidenced by an anomaly in the unit cell dimensions. The Pauli paramagnetism of La$_4$Ni$_3$O$_{10}$ changes character for $x = 0.05$ and temperature-dependent paramagnetism develops for $x \geq 0.25$. A Curie-Weiss regime appears to develop above 150 K.

No long-range magnetic order is detected by powder neutron diffraction. The introduction of the non-magnetic $Al^{3+}$ changes the $Ni^{3+}/Ni^{2+}$ ratio and is likely to block double-exchange pathways by means of -Ni-O-Al-O-Ni- fragments into the network of corner shared octahedra with the emergence of possible short-range order in ferromagnetic like islands.

## I.   INTRODUCTION

The Ruddlesden-Popper (RP) series of nickelates $La_{n+1}Ni_nO_{3n+1}$ attracts interest owing to their layer-like character, features resembling high-$T_c$ superconductors and due to their physical properties [1-4]. The RP$n$ crystal structure consists of $n$ consecutive perovskite layers, $(LaNiO_3)_n$, alternating with rock salt-like layers of LaO, $i.e.$ $(LaO)(LaNiO_3)_n$ [5]. The presence of half a rock-salt-like layer influences the physical properties and the number of the perovskite blocks ($n$) play an important role due to their two-dimensional like features [6-8]. With increasing $n$, the electrical behavior changes from insulating to metallic [9, 10]. Their structures are closely related, though with different symmetries where $La_2NiO_4$ is tetragonal [11], $La_3Ni_2O_7$ orthorhombic [12], and $La_4Ni_3O_{10}$ is slightly monoclinically distorted (pseudo-orthorhombic) [2, 4, 13] at room temperature. Oxygen non-stoichiometry (surplus) is quite extensive for the $n = 1$ member with additional O-atoms confined in the rock-salt-like layers [14].

Earlier we reported on crystal structure and phase transitions for $La_4Co_3O_{10}$ and $La_4Ni_3O_{10}$ and their solid solutions [13, 15-18]. In high-resolution synchrotron powder diffraction patterns, the monoclinic distortion is manifested by peak broadening, and in some cases, peak splitting. This $P2_1/a$ structure for $La_4Ni_3O_{10}$ was recently evaluated by DFT and found to be close in energy to variants with symmetries described in space groups $Bmab$, $Pcab$ as well in a smaller volume $P2_1/a$ structure ($P2_1/a$-II) [2, 19-20], likewise supported by single crystal diffraction studies [21]. Puggioni and Rondinelli found that the larger Z = 4 $P2_1/a$ structure transformed to the Z = 2 $P2_1/a$-II structure on coordinate relaxation [2]. In all cases the differences in terms of atomic displacements are small. We note that the atomic arrangement of $La_4Ni_3O_{10}$ is well approximated by the $Bmab$ B-centered orthorhombic model. Pristine $La_4Ni_3O_{10}$ is a Pauli-paramagnetic metal. An anomaly in the resistivity at ≈ 140 K is discussed in terms of a metal-to-metal transition (M-T-M) and/or charge density waves (CDW) [2, 9, 19, 22]. Further, recent experimental studies of the Fermi surface of $La_4Ni_3O_{10}$ using angle-resolved photoemission spectroscopy (ARPES) revealed a pseudo-gap at 120 - 150 K [23].

The average Ni-oxidation state increases for larger $n$-values $La_2(Ni^{2+})_1O_4$, $La_3(Ni^{2+})_1(Ni^{3+})_1O_7$, and $La_4(Ni^{2+})_1(Ni^{3+})_2O_{10}$. Further oxidation may take place by intercalated O-anions in the rock salt layers, whereas reduction gives random or ordered O-vacancies and may result in unusually low average oxidation state (+1.33 in $La_4Ni_3O_8$) [24, 25]. For these RP$n$ compounds, the physical properties are closely linked to their layer-like structure and dimensionality effects [26-29]. A rich number of phenomena may occur at low temperatures, like electronic and magnetic transitions, colossal magnetoresistance (MR), and long-range magnetic order [30-32]. To expand our insights, we currently explore the effect of substitution of non-magnetic $Al^{3+}$ into the Ni-sublattices of $La_4Ni_3O_{10}$. In general, the incorporation of chemical substitutions into $La_4Ni_3O_{10}$ is challenging, whether being substitution at the lanthanum sites with alkaline-earth cations, or at the Ni sites by transition-metal cations such as Mn, Co and Cu [13, 33-35]. Correspondingly, attention must be paid to avoid partial phase decomposition.

Importantly the fixed +3 oxidation state of the Al-substituent will provide electron doping that reduces the average oxidation state of nickel in $La_4Ni_3O_{10}$ while keeping the O-stoichiometry unchanged. Hence, the Al-substitution will affect (in ionic terms) the $Ni^{3+}/Ni^{2+}$ ratio and thus the exchange interactions. To minimize any potential effect of oxygen non-stoichiometry on the measured properties, all samples were heat-treated at T, $pO_2$ -conditions recognized to give stoichiometric $La_4Ni_3O_{10}$, which also ought to secure completely filled O-sites for the lower valent Ni-compositions resulting from Al-substitution.

We address several open questions: i) will Al-substitution trigger a metal-to-insulator (MIT) transition as function of temperature and/or composition? ii) will the CDW transition of $La_4Ni_3O_{10}$ disappear or change character? iii) will the Pauli paramagnetism transform into Curie-Weiss like paramagnetism with localized moments and long-range magnetic order emerging at low temperatures? iv) will we observe positive or negative magnetoresistance and how does this correlate with Al content and electrical resistivity? v) will changes in the formal $Ni^{3+}/Ni^{2+}$ ratio give rise to ferromagnetic like interactions via a double exchange mechanism? Herein we provide extensive new insights into structural, transport, magnetic and MR properties of $La_4Ni_{3-x}Al_xO_{10}$ ($0.00 \leq x \leq 1.00$) as studied by powder X-ray (PXRD) and powder neutron diffraction (PND), magnetization and resistivity studies. The atomic arrangement is described in space group $P2_1/a$ (Z = 4); partly approximated in *Bmab*. The experimental data are supported by first-principles DFT modelling with focus on electronic structure, bonding, oxidation states, magnetism and stability.

## II. METHODS

### A. Experimental methods

Samples of $La_4Ni_{3-x}Al_xO_{10}$ ($x$ = 0.00 - 1.00) were synthesized according to the citric acid method [13, 15-18]. The starting reactants were: $La_2O_3$ (99.99 %, Molycorp), $Ni(CH_3COO)_2 \times 4H_2O$ ($\geq$ 99 %, Sigma Aldrich), $Al(NO_2)_3 \times 9H_2O$ (98 %, Sigma Aldrich), $C_3H_4(OH)(COOH)_3 \times H_2O$ (98 %, Sigma Aldrich) and $HNO_3$ (min. 65 %, AnalaR NORMAPUR, VWR). Prior to use $La_2O_3$ was heated at 1273 K in air for 12 h and cooled to room temperature in a desiccator. Exact formula weights of the Ni- and Al-salts were determined gravimetrically, and standard salt solutions with accurate molar concentrations were prepared. $La_2O_3$ was dissolved in 6 M $HNO_3$ at 373 K under constant stirring before being mixed with thet standard Ni- and Al-solutions, according to the targeted stoichiometry. Subsequently, citric acid monohydrate was added in excess during stirring and heating of the entire solution. Nitrous gas species and water were boiled off until a syrup-like solution was formed, which was dried at 453 K overnight. The obtained crust was calcined at 673 K for 4 h and 1073 K for 1 h in static air, to remove organic matter. The resulting powders were ground, pelletized, and sintered in an alumina boat at high temperatures in air for 2 days before cooled to room temperature in the furnace. The chosen sintering temperature depended on the $Al^{3+}$ content. For higher $Al^{3+}$ contents the temperature required to stabilize a phase pure product increased. For $x$ = 0.00, sintering temperature was 1273 K, while 1473 K for $x$ = 1.00. Further, repeated sintering after one intermediate crushing was done to enhance sample homogeneity. An overview of all nominal compositions prepared are given in Supplemental Material Table S1.

The oxygen content of the prepared samples was determined by cerimetric titrations and thermogravimetric analysis (TGA). Cerimetric titration was performed by dissolving Mohr salt, $(NH_4)_2Fe(SO_4)_2 \cdot 6H_2O$ (99 %, Sigma-Aldrich), and a few mg of the sample in 1 M HCl in inert atmosphere of Ar (5N, Aga). The solution was then titrated with 0.1 M $Ce(SO_4)_2$ (volumetric, Fluka) under inert atmosphere of flowing Ar. The exact formula weight of the Mohr salt was determined gravimetrically using three parallel measurements. The exact concentration of the $Ce(SO_4)_2$ solution was determined separately by titrating with Mohr salt. TGA was performed in $N_2$ (5N Aga), $O_2$ (5N Aga), and $O_2/N_2$-atmospheres with a Netzsch STA 449 F1 Jupiter. Heating and cooling cycles were set at 10 °C/min. Background corrections were measured with empty sample holders under the same experimental conditions to eliminate any effects from the holders and the atmosphere.

Phase content and unit cell dimensions were analyzed by means of PXRD data collected at room temperature on a Bruker D8 Discover with a Lynxeye detector and Cu-$K_{\alpha 1}$ radiation [$\lambda$ = 1.5406 Å, Ge(111) Johansson monochromator]. The XRD data collected between 20° and 100° in steps of 0.02° were analyzed by Rietveld refinements using the TOPAS v5.0 software [36]. The background was modelled using a Chebyshev polynomial function and a Double-Voigt function to fit peak shapes. The initial structural parameters were taken from previously reported data (*Bmab, Pcab,* monoclinic-I $P2_1/a$ and monoclinic-II $P2_1/a$) for $La_4Ni_3O_{10}$ [2, 9, 13, 20]. Overall isotropic temperature factors for each element were refined. The occupation factors were fixed based on nominal sample stoichiometry, assuming that Al cations are randomly distributed over the Ni-sites and that all oxygen sites were fully occupied, *i.e.* having a fully stoichiometric RP3 phase. Absorption correction was included in the refinement.

PND data were collected at 12 K (Displex cryostat) for $La_4Ni_{2.50}Al_{0.50}O_{10}$ and $La_4Ni_2AlO_{10}$ with the high-resolution two-axis powder diffractometer PUS at the JEEP II reactor, Kjeller, Norway using monochromatic neutrons with $\lambda$ = 1.5558 Å [37]. The data were rebinned into steps of $\Delta 2\theta$ = 0.05° for the range $2\theta$ = 10° to 130°. The FULLPROF suite of programs [38] was used for Rietveld refinements, evaluating different proposed modifications of $La_4Ni_3O_{10}$. The following parameters were used, linear background corrections based on 20 background points (between 10 and 130°), the pseudo-Voigt profile function, overall isotropic temperature factors for La, Ni, Al and O were refined and scattering lengths La-0.824, Ni-1.03, Al-0.345 and O-0.580 fm were taken from the program library.

Low-temperature synchrotron PXRD data were measured at BM01 (Swiss-Norwegian Beamlines, SNBL), ESRF, Grenoble, France, at a wavelength of $\lambda$ = 0.71490 Å and using a 2D PILATUS2M detector [39]. Carefully ground powder was filled into a 0.3 mm diameter borosilicate capillary and cooled in an Oxford Cryostream 700+ nitrogen blower. XRD patterns were collected between 100 and 300 K at a detector distance of 300 mm. The 2D images were integrated using the SNBL Bubble software, with rebinned 1D diffraction data for $0 < 2\theta <$ 34.8° and step size 0.01° [39]. The patterns were simultaneously refined in a surface refinement using the TOPAS V5 software [36], with unit cell parameters being refined for each scan. Stephens model for anisotropic peak broadening was refined to all scans simultaneously [40].

A Quantum Design Physical Property Measurement System (QD-PPMS) was used for property characterization. Electrical resistivity was measured on bar-shaped bulk pellets, using in-line four-point probe contact geometry and a dc current ranging from 1 to 2 mA, for temperatures between 4 and 300 K. Magnetic properties were measured for samples held in gelatin capsules during warming cycles after at both zero field cooled (ZFC) and field cooled (FC) conditions over the temperature range 4 - 300 K. A persistent applied magnetic field of 0.2 T was used during ZFC/FC measurements. Magnetization was measured at 4 K between fields of ±9 T. Magnetoresistance (MR) curves were measured for fields up to 9 T at 4 K.

## B. Computational methods

Structural optimizations were performed using the Vienna ab-initio Simulation Package (VASP) [41]. We used a 7 × 7 × 3 Monkhorst-Pack k-point mesh [42] and a plane wave cut-off energy of 700 eV. Atoms were relaxed until the force on each atom was < 0.1 meV/Å. Perdew-Burke-Ernzerhof generalized gradient approximation (GGA) functional [43] was used. Owing to earlier experience with VASP being unable to successfully determine the ground-state magnetic ordering for RP nickelates [25], electronic and magnetic properties of $La_4Ni_{3-x}Al_xO_{10}$ ($x$ = 0.25, 0.50, 1.00) were studied by the full-potential linearized augmented plane wave (FP-LAPW) method (DFT) as implemented in the WIEN2k code [44]. The standard GGA exchange-correlation potential within the PBE-scheme was used. To describe the $3d$ electron-electron repulsion of the nickel cations, the GGA+$U$ method was used in the fully localized limit' (FLL). $U_{eff} = U$-$J$ ($U$ and $J$ are on-site Coulomb repulsion and exchange interaction, respectively) was used, with $U_{eff}$ = 6 eV. The muffin-tin (MT) radii for La, Ni, Al, and O were chosen to be 2.38, 1.91, 1.77 and 1.62 a.u. respectively. Integrations in reciprocal space were performed using 108 spatial $k$-points in the irreducible wedge of the Brillouin zone.

## III. RESULTS

### A. Crystal structure

We initiated our investigations on the $La_4Ni_{3-x}Al_xO_{10}$ ($x$ = 0.00 - 1.00) system by considering the crystal structure of the samples. All $La_4Ni_{3-x}Al_xO_{10}$ samples were synthesized according to the citric acid method, with special attention to the low Al content samples to assure good homogeneity concerning the Al/Ni distribution. Characterization by PXRD showed phase pure RP3 samples for 0.00 ≤ $x$ ≤ 1.00. A possible oxygen non-stoichiometry in $La_4Ni_3O_{10}$ has been discussed by several authors, however in line with our earlier studies [13, 15, 18], we could not prove any significant departure in oxygen content from the ideal stoichiometry (standard deviations of TGA < 1 % and cerimetric titration ±0.02). Therefore, the oxygen content is considered as 10.00 for all synthesized samples ($x$ = 0.00 - 1.00). These findings suggested that the samples are stoichiometric and that the samples are well suited for electron doping studies of $La_4Ni_3O_{10}$.

The unit cell dimension of $La_4Ni_{3-x}Al_xO_{10}$ ($x$ = 0.00 - 1.00) changes smoothly with increasing Al-content according to Rietveld refinements with an overall volume contraction, Fig. 1. This is consistent with the smaller ionic radius of $Al^{3+}$ (0.535 Å) compared to $Ni^{3+}$ (0.56 Å) and $Ni^{2+}$ (0.69 Å) [45]. The degree of distortion in the ab-plane diminishes for higher Al-levels, Fig. 1 (upper panel). As discussed above, Al-substitution provides electron doping to the Ni-atoms. The Ni-average oxidation state is thereby lowered from +2.67 for $x$ = 0.00 to +2.60 for $x$ = 0.50 and +2.50 for $x$ = 1.00.

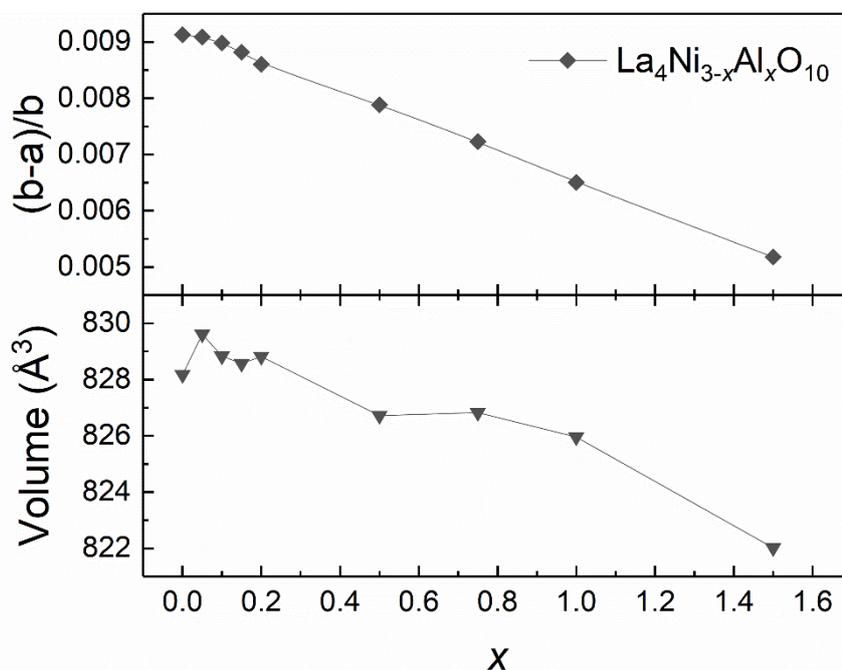

**FIG. 1.** Unit cell volume (lower panel) and degree of unit cell distortion in terms of $|(b - a)/a|$ (upper panel) versus Al-substitution ($x$) for $La_4Ni_{3-x}Al_xO_{10}$ from Rietveld refinements in space group *Bmab*.

The crystal structure of $La_4Ni_{3-x}Al_xO_{10}$ is revisited based on the recent DFT study that suggests a smaller monoclinic unit cell for the ground state [2]. To clarify which of the four theoretically reported structures that comply with our samples we evaluated home lab PXRD, synchrotron PXRD, as well as PND data. In the case of lower quality/lower resolution data (home lab; neutron) the differences in fits are small for these four structures. The departure from the orthorhombic metric is tiny with $\beta$ < 90.2° for a description in space group $P2_1/a$. Such minor monoclinic distortion was earlier identified by peak broadening in high-resolution powder synchrotron diffraction patterns of $La_4Co_3O_{10}$ and $La_4Ni_{3-x}Co_xO_{10}$ solid solution [13, 15, 16]. Hence, the peak width of (117) is a good measure of the degree of monoclinic distortion. This was currently evaluated by comparing peak profiles of (117/11-7), (200) and (020) for La$_4$Ni$_{3-x}$Al$_x$O$_{10}$, see Fig. 2, and it is evident that (117) peak has a broader profile than the (200) and (020) reflections. The (117) peak broadening is also evident for Al-substituted samples, hence La$_4$Ni$_{3-x}$Al$_x$O$_{10}$ exhibits a monoclinic distortion. A sketch of the crystal structure is shown in Supplemental Material Fig. S1.

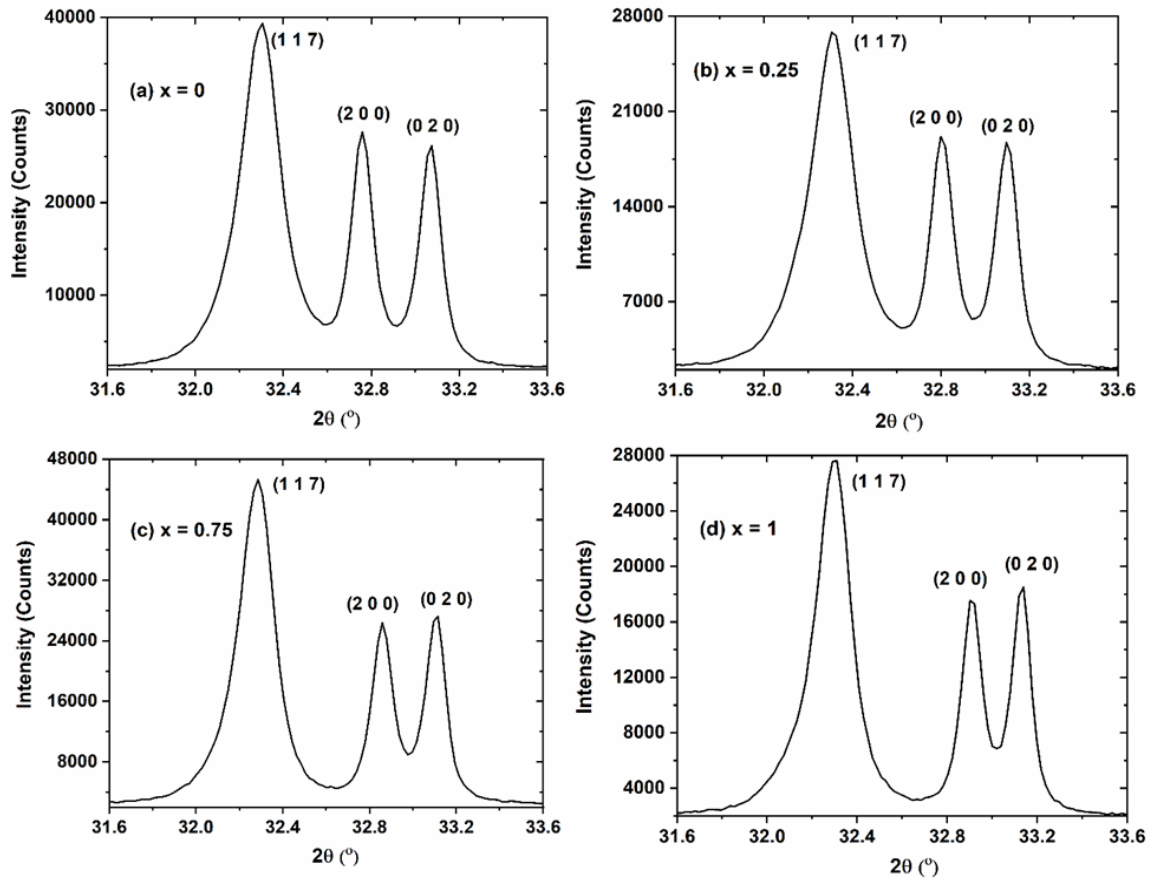

**FIG. 2.** (a) Rietveld fit to home laboratory powder X-ray diffraction data (left; λ = 1.5406 Å) showing a selected 2θ range where the enhanced peak width of (117)/(11-7) is evident compared to (020) and (200); space group $P2_1/a$. The (117) peak broadening is evident also for Al-substituted samples ((b), (c) and (d)).

To further evaluate our conclusion from diffraction data we turned to DFT and optimized the experimental descriptions of the orthorhombic and monoclinic variants of La$_4$Ni$_3$O$_{10}$. The total energy versus volume curves for the $P2_1/a$ and $Bmab$ models are shown in Fig. 3. Our calculations show that the monoclinic structure is more stable, however with a very small energy difference. Therefore, DFT calculations support our conclusion from analysis of diffraction data, *i.e.* that $La_4Ni_3O_{10}$ adopts the monoclinic structure ($P2_1/a$). By approximating the structure in space group *Bmab*, the number of coordinate variables is reduced from 48 to 13, which allows us to use home laboratory data for deriving variations in volume and atomic coordinates. Intensity profiles from the Rietveld fits are shown in Supplemental Material Fig. S2. When exploring the two possible models we note that the derived unit cell dimensions are very similar, with $La_4Ni_3O_{10}$ (*Bmab*) $a$ = 5.4167(8) Å, $b$ = 5.4667(5) Å, and $c$ = 27.967(4) Å, whereas for monoclinic ($P2_1/a$) $a$ = 5.4166(4) Å, $b$ = 5.4667(3) Å, and $c$ = 27.980(2) Å (and $\beta$ = 90.18°).

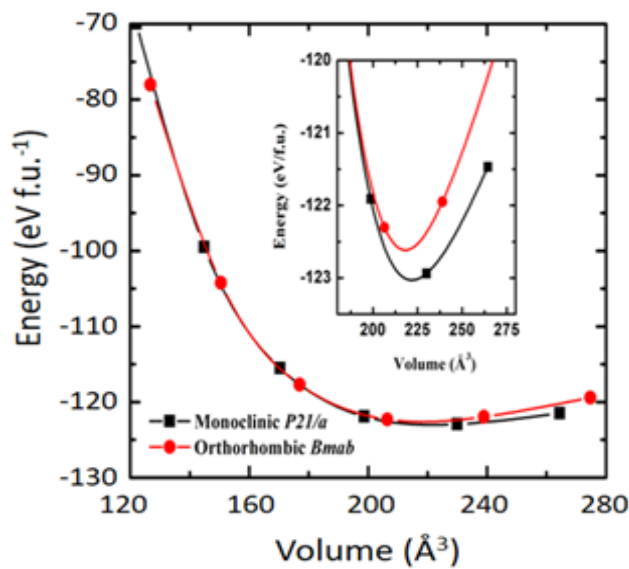

**FIG. 3.** Total energy versus volume (per f.u.) curves for *Bmab* and $P2_1/a$ structures of $La_4Ni_3O_{10}$, which show that the $P2_1/a$ structure indeed is most stable.

The Rietveld fit to the PND data for $La_4Ni_{2.50}Al_{0.50}O_{10}$ at 12 K is given in Fig. 4. The diffraction patterns of $La_4Ni_{2.50}Al_{0.50}O_{10}$ and $La_4Ni_2AlO_{10}$ are well described by their nuclear structure, hence long-range magnetic ordering is absent. The crystal structure contains two different types of Ni-environments located at the octahedra in the center of the perovskite block (Ni2) and at the outer part of the perovskite block (Ni1), respectively (see Supplemental Material Fig. S1). Based on the difference in scattering lengths between the Ni and Al atoms, we evaluated whether any preferences existed for Al substitution to either of the two Ni-sites. No such preference was detected. Hence, we conclude that Ni and Al are randomly distributed on the octahedral sites. For clarity, note that the number of non-equivalent Ni-sites is different for *Bmab* and $P2_1/a$, however there are still two different categories, Ni1 and Ni2 type environments. The Ni oxidation (charge) state and the Ni-O bonding for the non-equivalent Ni-cations is evaluated using the orbital-projected DOS, Born effective charge. In an ideal octahedral crystal field, the $d$-orbitals split into triply degenerate $t_{2g}$ and doubly degenerate $e_g$ states. In monoclinic fields, these degeneracies are lifted. However, in $La_4Ni_3O_{10}$ these levels are broadened into overlapping bands, and metallicity appears.

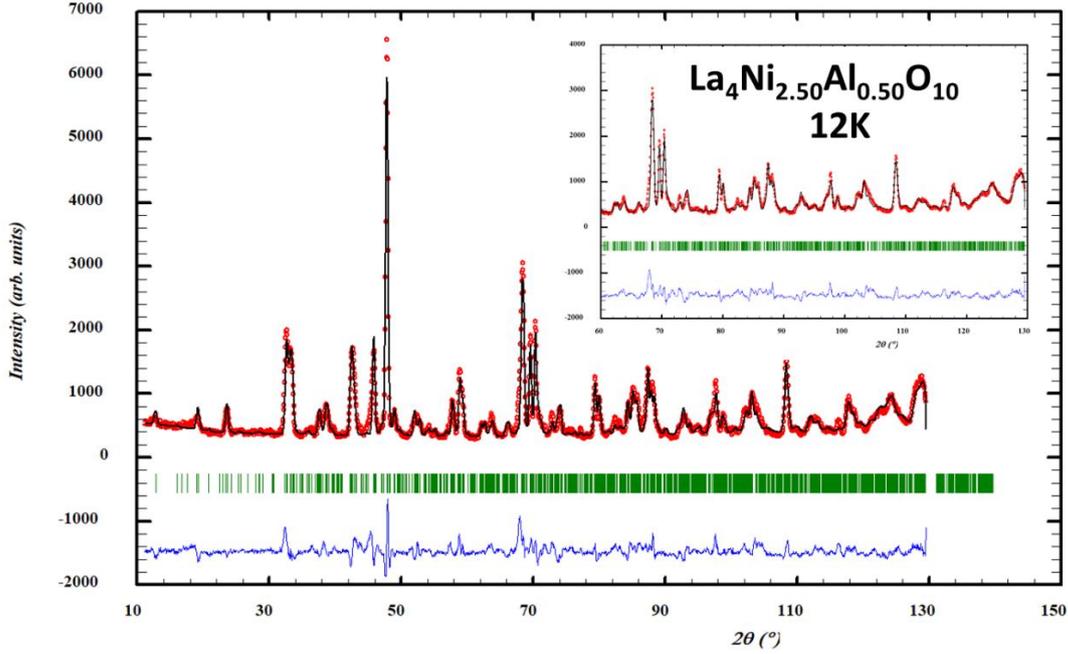

**FIG. 4.** Observed, calculated and difference intensity profile for Rietveld fit to powder neutron diffraction data of $La_4Ni_{2.50}Al_{0.50}O_{10}$, $\lambda = 1.5558$ Å. Space group $P2_1/a$.

The charge difference between the Ni1 and Ni2 muffin tin spheres for $La_4Ni_2AlO_{10}$ is about $0.06e$, which is minute compared to a possible nominal charge (ionicity) difference of $1e$ in case of charge ordering of Ni1 and Ni2. Figure 5 shows the orbital-projected DOS for Ni1 and Ni2 in a theoretical ferromagnetic ground state. For the Ni1 and Ni2 sites, both the majority and minority spins of the $d_{xy}$ and $d_{xz}$ orbitals are fully occupied. Similarly, the minority spin states of $e_g$ states are almost empty. The $d_{xy}$ orbitals for both types of Ni ions possess partial occupancy in the minority spin channel, whereas the majority channel is fully occupied. The integrated DOS gives $0.7e$ and $0.6e$ for Ni1 and Ni2, respectively. However, the extra contribution compared to the expected values of $0.5e$ for both ions (No oxidation state +2.5 on average) can be ascribed to hybridization between Ni $3d$ and O $2p$. These results suggest that Ni1 and Ni2 are in an average valence state between $Ni^{2+}$ and $Ni^{3+}$, and that no charge ordering is present. The average value of the diagonal components of the Born effective charges (BEC) tensor for La$_4$Ni$_2$AlO$_{10}$ is +2.4 and +2.7 for Ni1 and Ni2, respectively. The difference between the BECs of Ni1 and Ni2 of 0.3$e$ is minor as compared to a charge of 1e. Therefore, there is no indication for any Ni$^{3+}$ - Ni$^{2+}$ charge order with $d^7 + d^8$ species. The BEC for Al is $\approx$ +2.9, close to its nominal ionic value of +3.

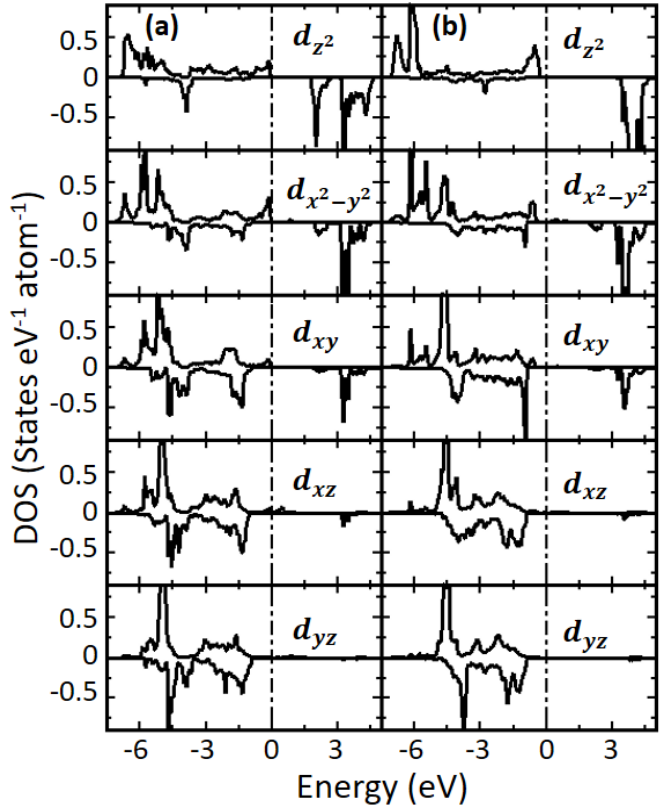

**FIG. 5.** Orbital projected DOS for Ni, (a) Ni1 and (b) Ni2 in ferromagnetic ground state for La$_4$Ni$_2$AlO$_{10}$ calculated by GGA+$U$ with $U_{eff}$ = 5 eV.

### B. Magnetic properties

Zero-field-cooled (ZFC) and field-cooled (FC) DC magnetization data (at 0.2 T persistent field) are shown for La$_4$Ni$_{3-x}$Al$_x$O$_{10}$ samples in Fig. 6. For La$_4$Ni$_3$O$_{10}$, the ZFC and FC curves show two regimes; below 100 K there is a weak tendency for localized paramagnetic behavior, whereas above 100 K almost temperature-independent Pauli paramagnetic behavior exists. This is qualitatively in agreement with previous reports [13, 18]. The M(T) data for $x$ = 0.02, 0.03, 0.05 and 0.15 (see Fig. S3) are qualitatively similar to those for La$_4$Ni$_3$O$_{10}$. We conclude that Pauli paramagnetic behavior prevails for $0 \leq x < 0.25$ [A weak feature at around 60 K in some sets of FC data is ascribed to minute amounts of adsorbed oxygen, owing to a minute leakage in the PPMS and the high surface area of the wet chemically prepared oxide powders (~ 1 μm particle size)].

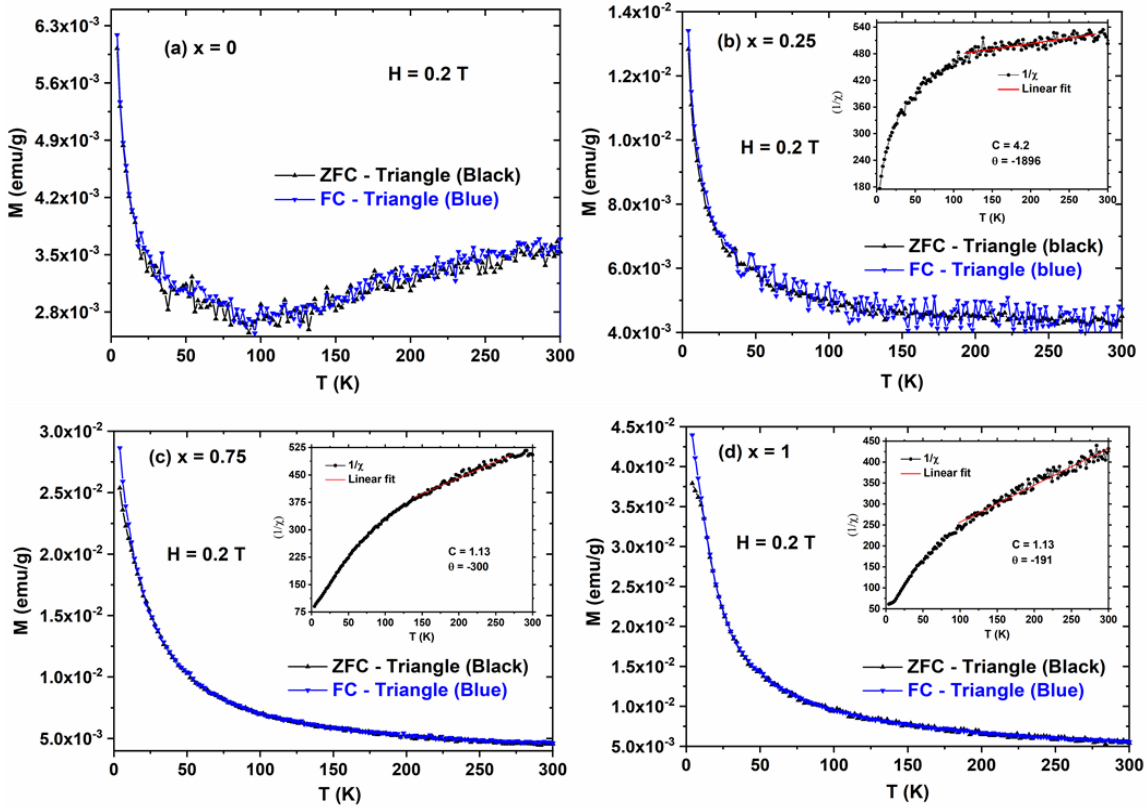

**FIG. 6.** M(T) magnetization data for $La_4Ni_{3-x}Al_xO_{10}$. Compositions ($x$) are given on the individual graphs (a-d). Data are measured in a field of 0.2 T. Inserts show inverse susceptibility versus temperature.

The magnetization M(T) for $x$ = 0.25, 0.75 and 1.00 is significantly higher than for Pauli paramagnetic samples at low Al contents, see Fig. 6. Their temperature dependences are much more pronounced, suggesting localized paramagnetic moments. However, there are no well-behaved Curie-Weiss regions in the inverse magnetic susceptibility data, see insets in Fig. 6. Nevertheless, a fit for the temperature range 150 - 300 K for $x$ = 0.75 gives the parameters $\mu_{eff}$ = 3.51 $\mu_B$ and $\theta$ = -315 K, indicative of antiferromagnetic spin interactions (for $x$ = 1.00; $\mu_{eff}$ = 3.31 $\mu_B$; $\theta$ = -205 K). At temperatures below 20 K, a bifurcation develops in the ZFC and FC magnetization curves for $x$ = 0.25 - 1.00. For $x$ = 0.00, the M(H) curve increases linearly with applied field consistent with Pauli paramagnetic behavior, Fig. 7(a). For the heavily Al-substituted samples $x$ = 0.25 - 1.00, Fig. 7(b), a small hysteresis develops on increasing Al-content. The magnetization at 4 K increases substantially with increasing Al-content. However, no samples reach saturation at the maximum field of 9 T.

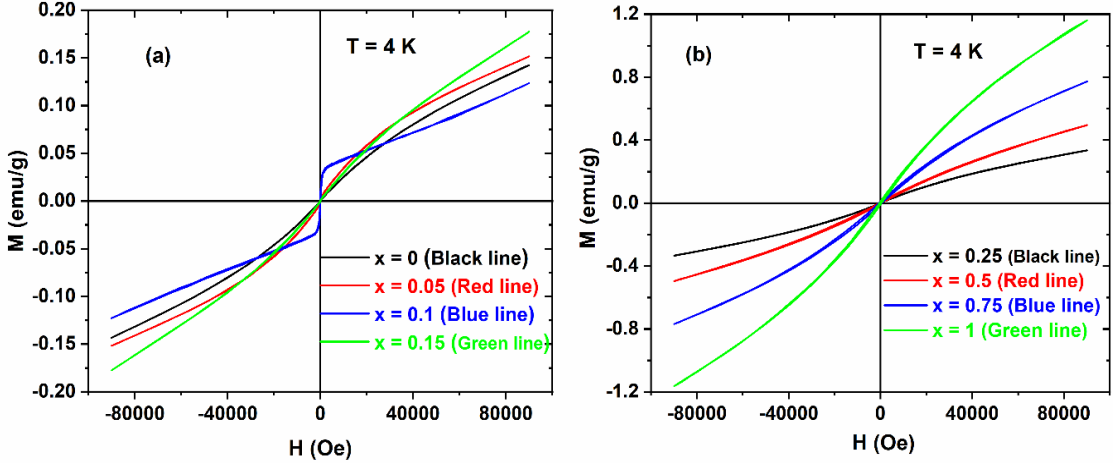

**FIG. 7.** M(H) data for $La_4Ni_{3-x}Al_xO_{10}$ at 4 K. Compositions ($x$) are given on the Figures (a) low Al-contents; (b) higher Al-contents.

The introduced non-magnetic (diamagnetic) $Al^{3+}$ ions will themselves not contribute directly to the observed magnetization. However, in an ionic picture, the substitution of $Al^{3+}$ for $Ni^{3+}$ will as such reduce the paramagnetic moment per formula unit, but notably, the introduction of $Al^{3+}$ imposes indirectly a higher content of $Ni^{2+}$ with S = 1 [relative to S = ½ for $Ni^{3+}$] and hence a higher moment. Double exchange is likely to exist between the $Ni^{2+}$ and $Ni^{3+}$ cations and mediated by O-anions in the semiconducting regime of the solid solution (see below). Note, these exchange pathways may partly be broken by non-magnetic $Al^{3+}$ cations. Hence, we foresee the possibility of (sub)nanometer-sized ranges of ferromagnetically interacting Ni-cations whereas 3D long-range ordering is prohibited. Interestingly, we note that ferromagnetic-like behavior is currently observed for $x$ = 0.10, see Fig. 7(a), with a small, yet distinct jump in the magnetization close to zero applied external field. We interpret the tiny hysteresis seen in Fig. 7(a) as an indication of small ferromagnetic like islands in the material.

### C. Resistivity – electronic and structural transitions

The temperature dependence of the electrical resistivity ($\rho$) of $La_4Ni_{3-x}Al_xO_{10}$ is shown in Fig. 8. The resistivity curves for $x$ = 0.00 and 0.02 in Figures 8(a) and 8(b) show two metallic regimes that are separated by a transition at around 140 K. For slightly higher Al-contents, $x = 0.05$, the metallic range at low temperatures changes character towards non-metallic, Fig. 8(c). With further increase of Al-contents, the transition regime shifted to the high temperature side and visible up to $x = 0.15$ ($\approx$ 160 K), Fig. 8(e). When the Al-content above $x = 0.15$ level, the high temperature metallic region vanishes and for $x = 0.25$ and 0.50 semiconducting behavior is observed, Figures 8(f) and 8(g). For $x > 0.50$, the samples become increasingly more resistive [Figures 8(h) and 8(i)]. The observed variation in resistivity is quite extraordinary. From diffraction and DFT analysis, we know that the two non-equivalent Ni1 and Ni2 atoms behave very similarly with respect to bonding, ionicity and charge state. Owing to our use of wet chemical synthesis routes and annealing efforts to assure homogeneous Al-distribution in the crystallites, these drastic changes are considered induced by Ni-sublattice dilution and electron doping.

We recently reported in detail on the physical properties of $La_4Ni_3O_{10}$ as a function of temperature and magnetic field [18]. Structurally, we then observed an anomaly in the unit cell parameters at the metal-to-metal (or the so-called CDW) transition around temperature of 140 K. Through heat capacity measurements, we demonstrated that this transition was of second-order, as also indicated by the variable temperature diffraction data. A most striking aspect of the composition dependent resistivity measurements in Fig. 8, is the extent to which signatures related to the M-T-M/CDW transition is able to maintain themselves in the face of substitutional disorder until levels of $x = 0.15$. This broad compositional range where the transition is preserved and enhanced, makes the Al-substituted system the most exciting one to study for exploring effects of doping and chemical pressure on electronic properties of the 2D $La_4Ni_3O_{10}$ material. At higher Al-contents, the M-T-M/CDW behavior is completely suppressed. We address two more important observations; (i) the temperature dependent changes of the resistivity above and below the transition; and (ii) the resistivity minimum with respect to the transition. For the high temperature region (above the transition), we observe that the linear T-dependence of metallic characters is maintained until $x = 0.15$, although the system gets relatively more resistive. The conduction regime changes to either higher-order or more complex power-law relations for $x > 0.15$. For $x < 0.25$, the samples are still within a metallic regime.

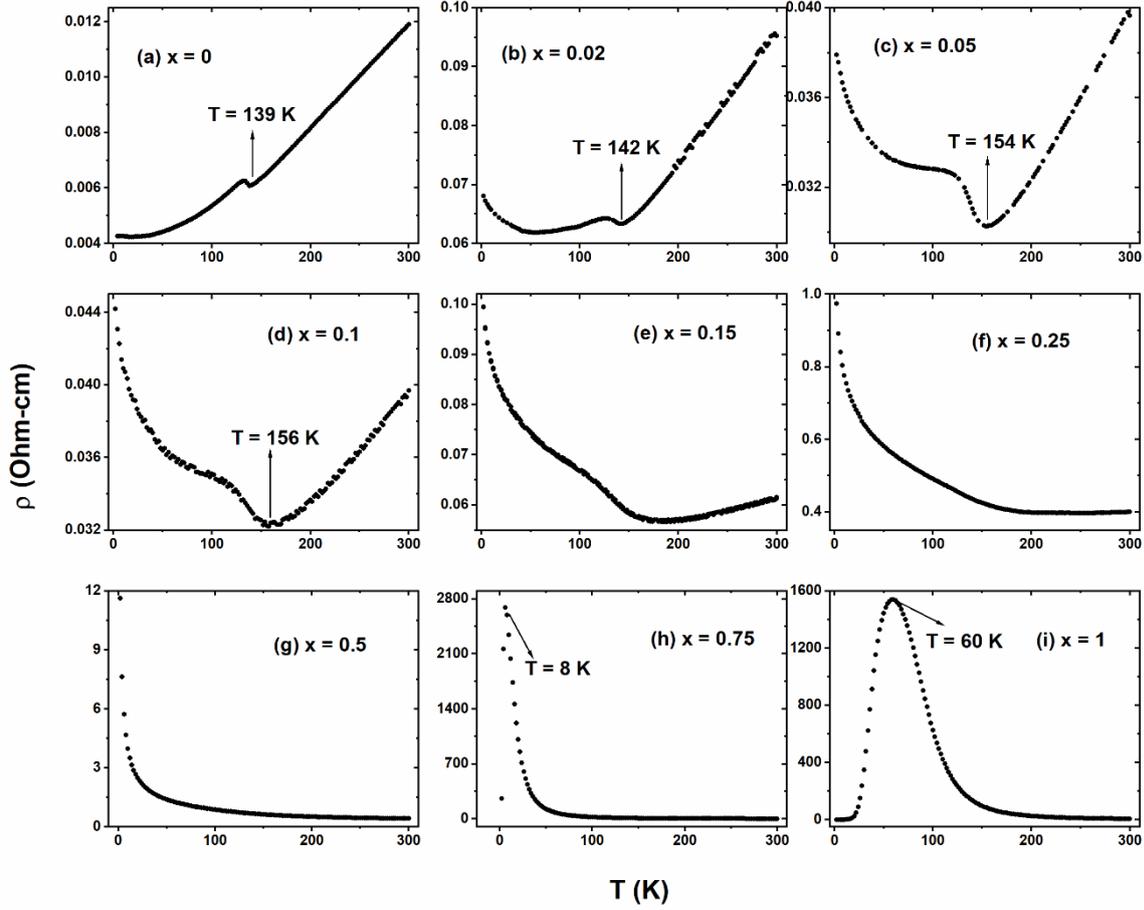

**FIG. 8.** Resistivity versus temperature for $La_4Ni_{3-x}Al_xO_{10}$. Al-substitution levels ($x$) are given on the individual graphs.

The most fascinating changes take place in the low temperature region (below the transition). We observe an immediate change for the resistivity minimum of $La_4Ni_3O_{10}$ at $T$ = 20 K shifting to 50 K for $x$ = 0.02, Fig. 8(b). For $La_4Ni_3O_{10}$ this minimum is reported to be due to electron-electron scattering events, generally showing $T^{-1/2}$ dependence. With gradual substitution of $Al^{3+}$ into $La_4Ni_3O_{10}$ ($x \geq 0.02$), we observe a composition dependent transformation from a metal-to-metal-like transition into an insulator-to-metal-like transition on increasing temperature (transition survives until $x$ = 0.15), and the onset temperature becomes the new MIT and varies slightly with small changes in Al-content, $T_{MIT} \approx 142, \approx 154, \approx 156$ and $\approx 160$ K for $x$ = 0.02, $x$ = 0.05, 0.10 and 0.15, respectively. Some 50 - 110 K below these onset temperatures we observe a broad hump in resistivity. An exciting aspect is the fact that increased substitution of Al results in an enhancement of this hump and a noticeable upturn in resistivity closely resembling what was reported for $Nd_4Ni_3O_{10}$ and $Pr_4Ni_3O_{10}$ (these have a compressed unit cell volume relative to $La_4Ni_3O_{10}$) [46, 47]. For $x$ =

0.25 and 0.5, the samples are in the semiconductor regime. When Al-content reaches $x = 0.75$ level, where we see the hints of the first temperature dependent MIT at the lowest temperatures ($\approx 8$ K) and this MIT moves to elevated temperatures ($T_{MIT} = 60$ K) for $x = 1.00$.

For $La_4Ni_3O_{10}$ it has been speculated whether the metal-to-metal transition involves closely related distorted structures [2, 18]. By means of high-resolution synchrotron powder diffraction, it was proved that the transition is connected with a structural anomaly, with $La_4Ni_3O_{10}$ being isostructural below and above the transition [13, 18]. The current detailed analysis of variable temperature synchrotron diffraction data for $x = 0.02$, shows a significant change in the *b*-axis (space group $P2_1/a$) at the metal-to-metal transition, Fig. 9(a). This structural transition correlates well with the anomaly in resistivity, Fig. 9(b). There is an excellent correlation between the resistivity and magnetization data shown by the metallic and Pauli paramagnetic regime for $La_4Ni_{3-x}Al_xO_{10}$ with $x < 0.25$. At higher Al-contents, a semiconducting/insulating state with localized moments occur.

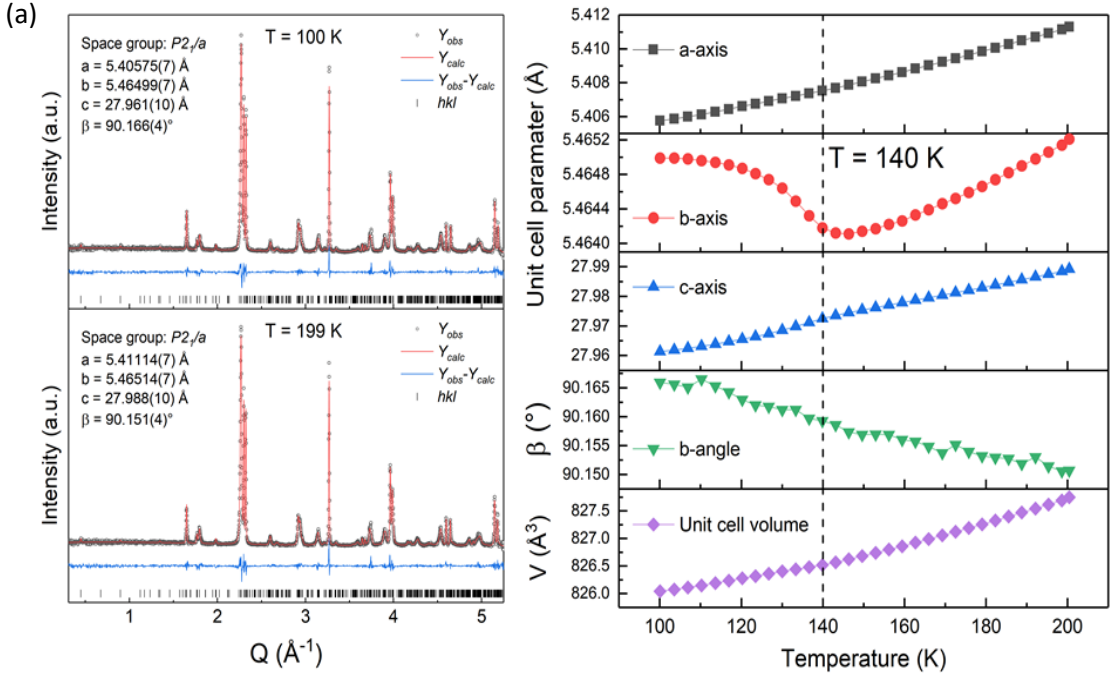

**FIG. 9.** (a) Rietveld fit to high resolution powder synchrotron diffraction data, space group $P2_1/a$ (left), and temperature dependence of unit cell dimensions for $x = 0.02$ (right).

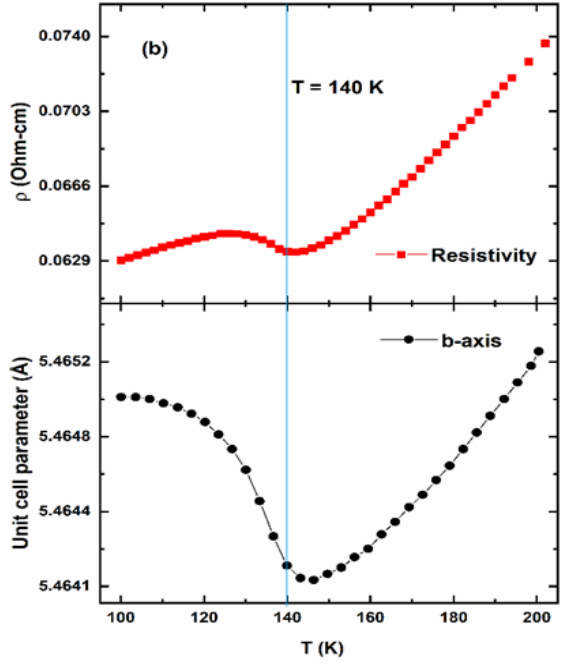

**FIG. 9.** (b) Temperature dependence of unit cell dimension and resistivity data for $x = 0.02$ showing an excellent correlation of structural anomaly associated with the metal-to-metal transition.

A localization of electronic charge is probably occurring due to both substitutional disorder and enhancement of anisotropy in conduction due to Al substitution. This results in increasing electron-electron scattering events as the Al content increases, until a composition dependent suppression of the M-T-M/CDW transition takes place. The introduction of $Al^{3+}$ introduces disorder and more anisotropy, possibly weak Anderson localization, which quickly becomes evident in the low temperature resistivity data since the substitution disrupts the conduction electron behavior, but without the resistivity jumping to high values. Since $Al^{3+}$ is non-magnetic, Pauli paramagnetic behavior will remain in the range where metallic conductivity is observed, up to $x < 0.25$. The 2D nature of the atomic arrangement of $La_4Ni_{3-x}Al_xO_{10}$ may possibly be mimicked by nickelate superlattices. These are foreseen to exhibit metal-insulator phenomena as outlined by Chaloupka and Khaliullin [48] and Park *et al* [49]. Recently, supporting literature towards such possible superlattice constructions is demonstrated for single crystals of $Pr_4Ni_3O_{10}$ [50]. Interestingly, they reported that the in-plane ($\rho_\parallel$) resistivity below the M-T-M/CWD transition is enhanced by a factor six at the lowest temperatures relative to the out-of-plane ($\rho\perp$) resistivity, indicating extreme localization of charge carriers within the quasi-2D conduction layers.

## D. Magnetoresistance

For $La_4Ni_{3-x}Al_xO_{10}$ we observe both positive and negative magnetoresistance, dependent on the Al-substitution level, Fig. 10(a). The magnetoresistance (MR) is shown for $x$ = 0.25, 0.50 and 0.75 at 4K in the field range 0 < H < 9 T in Fig. 10(b); MR % = {[$\rho(H)$ - $\rho(0)$]/$\rho(0)$} × 100, where $\rho(H)$ and $\rho(0)$ are resistivities for the given applied magnetic field H and zero field, respectively [51]. As far as susceptibility data is concerned, we see a shift from low substitution Pauli paramagnetic behavior transition into a localized paramagnet with possible short-range ferromagnetic ordering for $x$ = 0.75 and 1.00. Looking at it from spin-dependent transport aspect, we observe a substitution driven enhancement of the electronic non-uniformity (induces local magnetic moments) which in turn increases spin-dependent scattering resulting in negative MR effect from $x$ = 0.25 to 0.50, as seen previously in manganese perovskites [52, 53]. Similar to reported mechanisms for manganese perovskites, we observe an increase in disorder and anisotropy as more and more Al is substituted into $La_4Ni_3O_{10}$, and we note that both magnetic anisotropy and spin-polarized transport across grain boundaries (percolative transport) are sensitive to disorder and substitution and hence tend to affect MR and its sign directly.

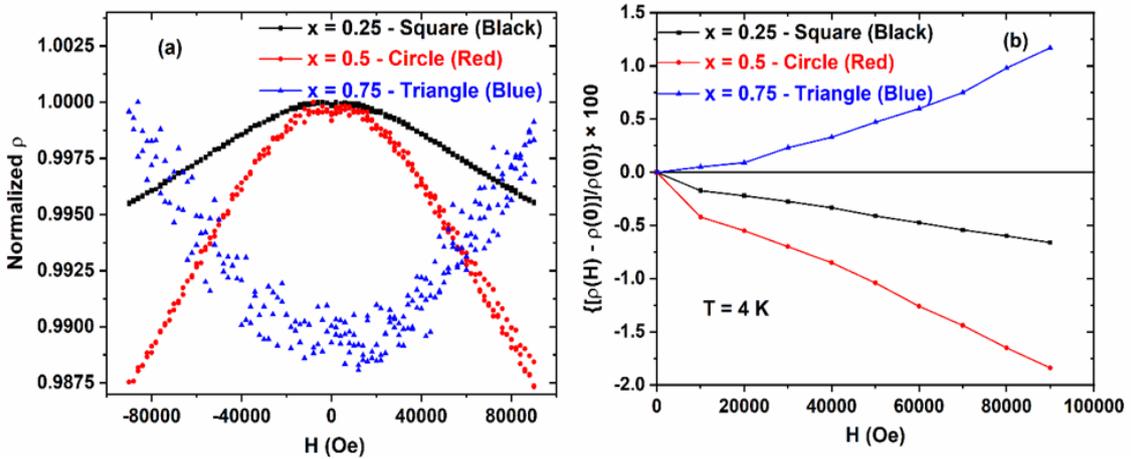

**FIG. 10.** (a) Normalized resistivity for $x$ = 0.25, 0.50 and 0.75 versus applied field at 4 K. (b) Variation in magnetoresistance for $x$ = 0.25, 0.50 and 0.75 versus magnetic field.

For $0.25 \leq x \leq 0.50$, as the magnetic field increases, the (negative) MR increases in magnitude and reaches a maximum of -2 % for $x$ = 0.50. The negative MR response is indicative here of antiferromagnetic interactions, which are short-range in nature. However, at the value of $x = 0.75$ of Al-substitution, there are fewer conduction pathways for those same majority spin carriers, and as a result, the MR effect is positive. The change in sign of the MR response indicates that the Al-ions block more and more conduction pathways ($Ni^{2+}$-O-Al-O-$Ni^{3+}$) with short-range ferromagnetic interactions and the samples become more and more resistive. Therefore, without the application of a magnetic field ($x > 0.50$), the spins of the prevailing ferromagnetic-like components (within nanometer-sized domains) are randomly oriented, and the transport is limited by spin scattering. Under an applied magnetic field, these ferromagnetic-like components are aligned, resulting in spin-dependent transport. The positive MR effect at higher $Al^{3+}$ contents correlates with the increase in the ferromagnetic component, as seen in the M(H) data at low temperatures, and furthermore this correlates with the change in the formal $Ni^{3+}/Ni^{2+}$ ratio upon $Al^{3+}$ substitution.

### E. Electronic structure

The recent study on $La_4Ni_3O_{10}$ shows that the electronic structure is qualitatively similar for the monoclinic and orthorhombic variants [2]. Hence, we performed our studies in the higher symmetric orthorhombic description since Al-substitution involves expansion into large supercells. The electronic band structure of $La_4Ni_3O_{10}$ (*Bmab*) is showed in Fig. 11(a), and exhibits features as reported in [2]. The γ bands with $d_{z^2}$ orbital characters are flattened in the vicinity of the Fermi level. This agrees well with the pseudo-gap obtained in the symmetrized energy distribution curves obtained at 24 K [23]. To clarify pressure effects on the electronic structure of $La_4Ni_3O_{10}$ caused by internal chemical pressure, a high-pressure (HP) situation with 5 % less unit cell volume is compared with the ambient pressure (AP) phase (ground state volume structure). This data is also relevant for isostructural compounds where the La-cation is exchanged with smaller rare earth cations, and thus comparisons with $Nd_4Ni_3O_{10}$ and $Pr_4Ni_3O_{10}$. Note that the introduction of $Al^{3+}$ gives a volume contraction, though less than 1 %, see Fig. 1.

The band dispersions along the M-Y line are compared in Fig. 11(b). We note that the bands become more dispersed for the HP situation. This is because of the pressure on the system, which can broaden the bandwidth by enhancing the band overlap. This leads to a decrease in resistivity with increased pressure, which is consistent with a decrease in resistivity of $La_4Ni_3O_{10}$ with increasing pressure. Hence the $La_4Ni_3O_{10}$ remains metallic under pressure with an increase in conductivity because of the increase in bandwidth (so the carrier effective mass) as compared to the ambient pressure case. These results are consistent with the pressure versus resistivity data obtained by Wu *et al.* [54].

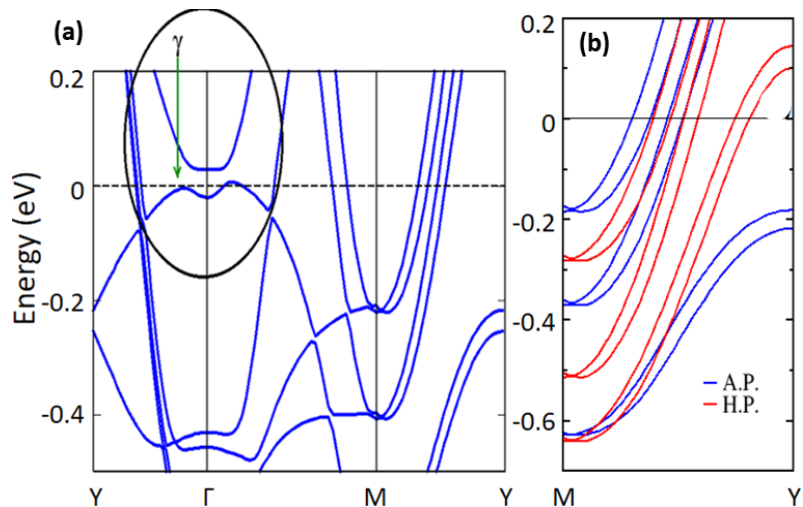

**FIG. 11.** (a) Electronic band structure of $La_4Ni_3O_{10}$ in the *Bmab* space group; (b) The electronic band dispersion along M-Y for the ground state at ambient pressure (A.P.) and at high pressure (H.P.) with a 5 % smaller unit cell volume.

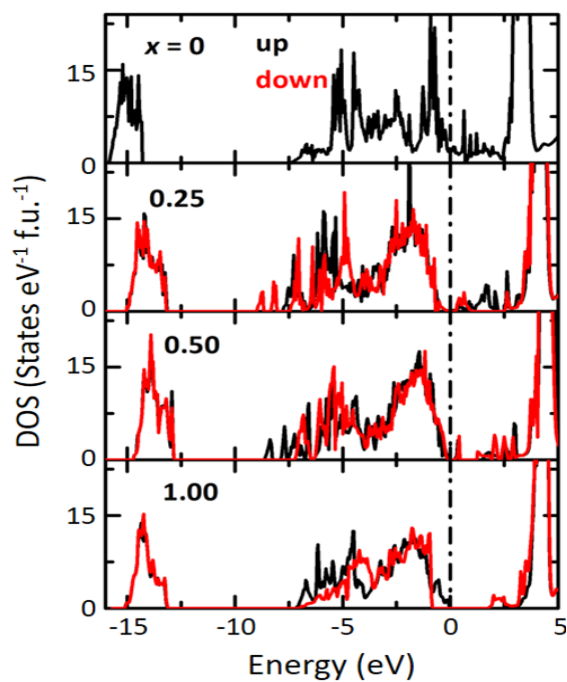

**FIG. 12.** Total DOS for $x = 0.00$, 0.25, 0.50 and 1.00 calculated with GGA+$U$, $U_{eff} = 6$ eV. The black and red lines indicate the up and down spin channels, respectively.

The calculated total DOS for La$_4$Ni$_{3-x}$Al$_x$O$_{10}$, $x = 0.00$, 0.25, 0.50 and 1.00, in the ground state magnetic configuration is given in Fig. 12. We show the highest occupied energy level in the valence band (VB), $E_F$, by the dotted line. In the case of $x = 0.00$, finite DOS values are present near $E_F$ (for both *Bmab* and *P2$_1$/a*). Hence, La$_4$Ni$_3$O$_{10}$ exhibits metallic character, consistent with resistivity measurements. For $x = 0.25$, 0.50, and 1.00, electrons start to localize, as seen from the reduced number of states at $E_F$. For $x = 0.25$, a finite bandgap of 0.28 eV opens between the top of the VB and the bottom of the conduction band (CB), *i.e.*, $x = 0.25$ shows semiconducting behavior, consistent with experiments. The bandgap evolution with increasing Al content is given in Table I.

**TABLE I.** Calculated ground state magnetic ordering, magnetic moments of Ni atoms, Ni-O bond length, Ni-O-Ni bond angle and band-gap for Al contents of $x = 0.00$, 0.25, 0.50 and 1.00.

| Al content ($x$) | Magnetic configuration | Ni1/Ni2 Magnetic moment ($\mu_B$) | Average Ni1-O/Ni2-O bond (Å) | Ni1 - O - Ni2 bond angles in degree | Band gap (eV) |
|---|---|---|---|---|---|
| 0.00 | Paramagnetic | - | 1.97/1.92 | 166.3 | Metal |
| 0.25 | Ferrimagnetic | 1.09/1.39 | 1.99/1.93 | 163.4 | 0.28 |
| 0.50 | Ferrimagnetic | 1.36/1.50 | 2.02/1.94 | 162.3 | 1.12 |
| 1.00 | Ferromagnetic | 1.43/1.72 | 2.04/1.96 | 161.4 | 1.81 |

Figure 13 shows the angular momentum projected DOS for the La, Ni and O atoms. The valence band of La$_4$Ni$_3$O$_{10}$ is primarily based on Ni 3*d* and O 2*p* admixtures. The negligibly small DOS contribution in VB from the La site indicates pronounced ionic bonding between La and O. The very similar topology of the DOS profile for the Ni sites supports the presence of an average valence state for all Ni atoms. The *s* and *p* states of Ni have a negligible contribution in VB in vicinity of $E_F$. Furthermore, Fig. 13 shows the partial DOS for the three types of oxygen atoms O1, O2, and O3 situated in the Ni1 layer, Ni2 layer, and La layer, respectively. The *d* states of Ni and *p* states of O are distributed over the range -7 eV to 4 eV. This implies that the Ni *d* and O 2*p* states are strongly hybridized and the bands cross the Fermi level resulting in metallicity. The O *2s* states are localized around -14 eV. Inclusion of parameter *U* is necessary for stabilizing a ferrimagnetic ground state relative to a ferromagnetic state for both $x = 0.25$, 0.50 and 1. For $U = 0$ a metallic ferromagnetic state is favored. The results presented here refer to $U = 6$ eV and $J = 1$ eV. The calculated values for magnetic moments and the type of ground state ordering is given in Table I. DFT calculations are consistent with Pauli paramagnetic ground state observed for $La_4Ni_3O_{10}$. For $x = 0.25$ and $0.50$ a ferrimagnetic configuration is calculated as ground state, whereas $x = 1.00$ stabilizes in a ferromagnetic configuration. This is partly in agreement with experiments.

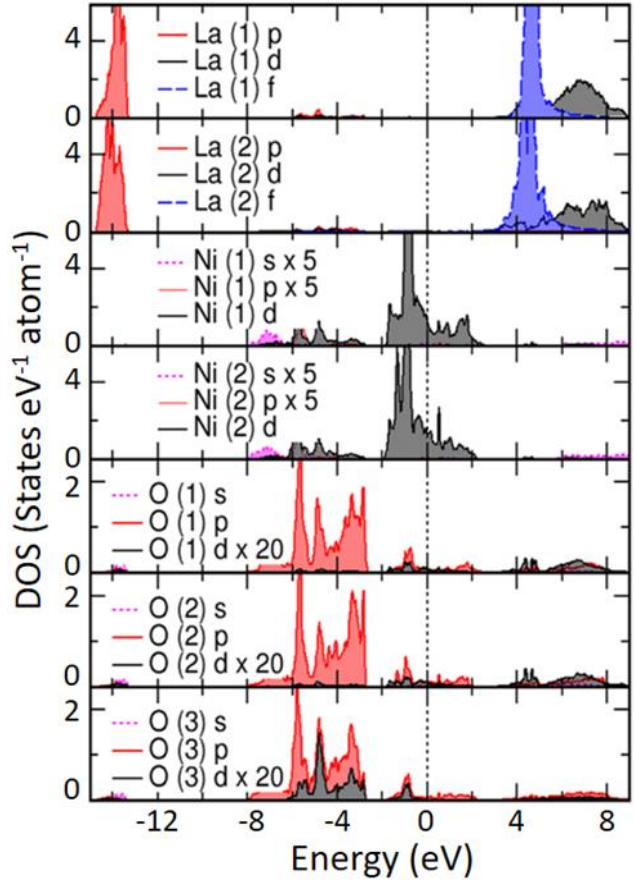

**FIG. 13.** Angular momentum projected DOS for the constituent atoms of $La_4Ni_3O_{10}$ calculated with GGA+$U$ for $U_{eff} = 6$ eV.

We do observe increased ferromagnetic interactions on increasing Al-levels in M(T) and M(H) data, however, there is no indication for any long-range magnetic order. We note that the ferrimagnetic configuration is essential in order to model an appropriate semiconducting ground state for $x = 0.25$ and $0.50$ that in turn will be consistent with the resistivity data. The calculated moment of some 1.4 $\mu_B$ (Table I) is less than the 2S value derived from the Curie Weiss fit, however, the latter region is not well defined in the M(T) data, see above Fig. 6.

# IV. Discussion

The metallic properties of $La_4Ni_3O_{10}$ are due to mixing of oxygen 2$p$ and nickel 3$d$ orbitals. For $La_4Ni_{3-x}Al_xO_{10}$, the $e_g$ electrons and the average valency of Ni ions ($Ni^{+2.67}$) provide the basis for electronic conductivity and magnetic exchange via O-atoms. Homogeneous Al-substitution in $La_4Ni_{3-x}Al_xO_{10}$ has at least three effects on the transport properties; first by blocking pathways for electron jumping between Ni-atoms located in the center of corner shared $NiO_6$-octahedra, second by enhancing the formal $Ni^{2+}$ content of the mixed-valence state of nickel and thereby increasing Ni-O bond distances, and third by opening up a gap at the Fermi level.

The Ni-O-Ni interactions are strongly perturbed by randomly distributed $Al^{3+}$ cations in the structure. The absence of magnetic interactions connected to these $Al^{3+}$ ions gives rise to Ni-O-Al-O-Ni structural fragments that weaken the magnetic exchange pathways and reduces the number of charge carriers that possibly participate in $Ni^{2+}$-O-$Ni^{3+}$ double-exchange. Experiments show that the metallic behavior is completely lost at moderate levels of Al substitution. Hence, the Al-richer $La_4Ni_{3-x}Al_xO_{10}$ samples are semiconducting/insulating with electron hopping appearing as a likely conduction mechanism favoring double exchange. Meaning, a changeover from metallic to semiconducting to insulating behavior on Al-substitution is seen. In terms of electronic structure, as described by DFT the main effect of Al-substitution is to lower the DOS at the Fermi level with the consequence that a band-gap is opening which increases in magnitude on increased Al-substitution.

The electronic conductivity and the magnetic properties are extraordinarily dependent on the Al-substitution level. With gradual substitution of $Al^{3+}$ into $La_4Ni_3O_{10}$, we tend to align the low temperature insulating phase to the nearest M-T-M/CDW transition and the onset temperature becomes the new MIT until $x < 0.25$ (composition dependent MIT). For $x = 0.25$ to 0.50, the material is in the paramagnetic semiconductor regime till the lowest of temperatures (as negative MR clearly supports at 4 K). Here if there are any short-range ordering, it should be antiferromagnetic. When the Al content hit $x = 0.75$ level, where there are hints of the first temperature dependent MIT at the lowest temperatures (MR becomes positive) and this MIT moves to elevated temperatures ($T_{MIT} = 60$ K). At these compositions, the system is dominant (at low temperature) ferromagnetic-like ordering in the sea of a paramagnetic insulator.

The intriguing metal-to-metal transition of $La_4Ni_3O_{10}$ that attracts significant interests theoretically is retained at low Al-levels ($x = 0.15$), however at lower temperatures the electric properties change even for lower *x*-values character towards a less conducting state. To explore these intriguing details at an ultra-low substitution level, it is necessary to pay attention to material synthesis. Indeed, a wet chemical approach with gravimetrically analyzed precursors is a prerequisite for proper composition control as well for ensuring random Al-distributions at the cationic sites and throughout all crystallites being in the sub-micrometer range. The present synchrotron diffraction data provides evidence of a minor, yet significant change in some of the unit cell parameters at the metal-to-metal transition. Hence, there exists a distinct coupling albeit probably being weak.

Both experiments and DFT clearly indicate that the Ni-cations take an average valence state (an intermediate, average oxidation number) with no indication for any $Ni^{2+}$ - $Ni^{3+}$ charge ordering (which possibly would have been appropriate for an ionic material). The surroundings for the two types of Ni-atoms are different: the Ni-atoms in the central part of the perovskite block are surrounded by six Ni-atoms in neighboring octahedra, whereas the Ni-atoms in the outer part of the perovskite block has just five such neighboring octahedra. Statistically, some of these Ni-atoms will formally be $Ni^{2+}$ others $Ni^{3+}$, and one may envisage double exchange by electron hopping to be operative and provide (weak) ferromagnetic interactions. The substitution of Al for Ni will cause unavoidably break such exchange Ni-O-Ni pathways, *i.e.* by means of randomly inserted fragments of -Ni-O-Al-O-Ni-. Hence, local islands with (weak) ferromagnetic interactions may appear in the material. Similar behavior has previously been observed for alkaline earth-substituted manganites like $La_{0.7}Ca_{0.3}MnO_3$ [55]. We note furthermore a changeover from negative to positive magnetoresistance for $x > 0.50$. The enhanced (positive) MR effect at higher $Al^{3+}$ contents correlate with an increase in the (short-range) ferromagnetic component seen in M(H) data at low temperatures, and furthermore with the lowered $Ni^{3+}/Ni^{2+}$ ratio caused by $Al^{3+}$ substitution. The latter observation is interesting and points at a more general route to tune MR properties in $La_4Ni_3O_{10}$ and other Ruddlesden-Popper type oxides, from a technological perspective.

## V. Conclusion

In a metallic low-dimensional perovskite, such as $La_4Ni_3O_{10}$, electronic states are very sensitive to substitutional effects and the low temperature phase is the more-often-than-not the first to show such fluctuations in transport behavior. As the system here already shows anisotropic conduction to begin with Al-substitution would result in enhancement of both disorder and anisotropy possibly due to the preference of $Al^{3+}$ substituent ion reducing conduction pathways -Ni-O-Al-O-Ni-. Indeed, we see such direct effects on electronic conduction pathways even at $x = 0.02$ substitution levels, where the low temperature resistivity minima shift to elevated temperatures. With the introduction of slightly more $Al^{3+}$, $x = 0.05$, we observe the low temperature state to completely transform into semiconducting/insulating state. Whereas, the high temperature conduction, mostly assisted by phonons sees minor changes only. What is striking is the intermediate temperature region showing the transition (M-T-M/CDW) which not only enhances in intensity, but also the resistivity curve shape is strikingly similar to what is observed for $Nd/Pr_4Ni_3O_{10}$ counterparts [46, 47], indicating a contraction in the $La_4Ni_3O_{10}$ unit cell dimensions similar to the replacement of Nd/Pr for La. The underlying reason for the enhancement of the M-T-M/CDW transition for the moment remains unclear. Furthermore, we notice the onset temperature of M-T-M/CDW becomes the new MIT temperature from $x = 0.02$ up till $x < 0.25$, beyond which this transition is completely suppressed. However, at the same time the low temperature phase is truly semiconducting/insulating and is representative also in the MR which is negative for $x = 0.25$ pointing to the fact that, ferromagnetic domains/clusters that are formed due to Al-substitution ($Ni^{2+}$-O-Al-O-$Ni^{3+}$) show field dependent reduction in spin transport of majority spin carriers and is further lowered for $x = 0.50$. At $x = 0.50$, we observe a substitution dependent MIT from a coexisting low temperature paramagnetic insulating and high temperature metallic state into a purely paramagnetic insulating state, which shows up in MR curve as a sharpening of the negative MR effect. For $x = 0.75$, we again observe a very low temperature MIT and a sign change of MR from negative to positive with a ZFC-FC bifurcation possibly indicating the creation of short-range ferromagnetic ordering of the ferromagnetic domains/clusters that are formed due to Al-substitution ($Ni^{2+}$-O-Al-O-$Ni^{3+}$) in a paramagnetic insulating matrix. Finally, at $x = 1.00$ we observe the same ferromagnetic ordering strengthening as seen from magnetization curves taken at 4 K and an enhancement of the MIT temperature to around 60 K, possibly driven by polaron formation, as reported for manganese perovskites [52, 53, 56]. This subtle interplay of conduction properties with substitution and its influence to conduction pathways, where the substituent ions are non-magnetic makes $La_4Ni_{3-x}Al_xO_{10}$ a very exciting system to study in detail through transport, X-ray and neutron (inelastic and total scattering) diffraction studies. The very recent studies of single crystals of RP3 nickelates shed important light on charge and spin density waves, and such studies should be extended to include Al-substituted samples – if feasible for single crystal growth.

## ACKNOWLEDGMENTS

We thank Ingrid Marie Bergh Bakke for the synthesis of samples for powder neutron diffraction; SNBL@ESRF is gratefully acknowledged for providing beamtime at BM31 and excellent support; the Research Council of Norway (RCN) project NOFCO (RCN 221905) as well as the Norway – India collaborative projects TESFUN and INNOREM (SIU 10089 and RCN 275014) are acknowledged for financial support.
## REFERENCES

1. M. Greenblatt, Ruddlesden-Popper $Ln_{n+1}Ni_nO_{3n+1}$ nickelates: structure and properties, Curr. Opin. Solid State Mater. Sci., **2**, 174 (1997).
2. D. Puggioni and J. M. Rondinelli, Crystal structure stability and electronic properties of the layered nickelate $La_4Ni_3O_{10}$, Phys. Rev. B **97**, 115116 (2018).
3. F. Baiutti, G. Gregori, Y. E. Suyolcu, Y. Wang, G. Cristiani, W. Sigle, P. A. van Aken, G. Logvenov, and J. Maier, High-temperature superconductivity at the lanthanum cuprate/lanthanum–strontium nickelate interface, Nanoscale **10**, 8712 (2018).
4. A. Olafsen, H. Fjellvåg, and B. C. Hauback, Crystal Structure and Properties of $Nd_4Co_3O_{10+\delta}$ and $Nd_4Ni_3O_{10-\delta}$, J. Solid State Chem., **151**, 46 (2000).
5. S. N. Ruddlesden and P. Popper, The compound $Sr_3Ti_2O_7$ and its structure, Acta Crystallogr., **11**, 54 (1958).
6. A. S. Botana, V. Pardo, and M. R. Norman, Electron doped layered nickelates: Spanning the phase diagram of the cuprates, Phys. Rev. Mater., **1**, 021801 (2017).
7. R. Zhong, B. L. Winn, G. Gu, D. Reznik, and J. M. Tranquada, Evidence for a Nematic Phase in $La_{1.75}Sr_{0.25}NiO_4$, Phys. Rev. Lett., **118**, 177601 (2017).
8. R. Kajimoto, K. Ishizaka, H. Yoshizawa, and Y. Tokura, Spontaneous rearrangement of the checkerboard charge order to stripe order in $La_{1.5}Sr_{0.5}NiO_4$, Phys. Rev. B **67**, 014511 (2003).
9. Z. Zhang and M. Greenblatt, Synthesis, structure, and properties of $Ln_4Ni_3O_{10-\delta}$ (Ln= La, Pr, and Nd), J. Solid State Chem., **117**, 236 (1995).
10. V. I. Anisimov, D. Bukhvalov, and T. M. Rice, Electronic structure of possible nickelate analogs to the cuprates, Phys. Rev. B **59**, 7901 (1999).
11. T. Kajitani, S. Hosoya, M. Hirabayashi, T. Fukuda, and T. Onozuka, Crystal structure of tetragonal form of $La_2NiO_{4+x}$, J. Phys. Soc. Jpn., **58**, 3616 (1989).
26

# Supplemental Material information

**TABLE S1**. $La_4Ni_{3-x}Al_xO_{10}$ composition details ($x$ = 0.00 - 1.00) and sintering temperature required to stabilize each $x$ phase.

| Samples | $x = 0$ | $x = 0.02$ | $x = 0.03$ | $x = 0.05$ | $x = 0.1$ | $x = 0.15$ |
|---|---|---|---|---|---|---|
| Composition | $La_4Ni_3O_{10}$ | $La_4Ni_{2.98}Al_{0.02}O_{10}$ | $La_4Ni_{2.97}Al_{0.03}O_{10}$ | $La_4Ni_{2.95}Al_{0.05}O_{10}$ | $La_4Ni_{2.90}Al_{0.10}O_{10}$ | $La_4Ni_{2.85}Al_{0.15}O_{10}$ |
| Content of Aluminum (%Al) | 0 | 0.007% | 0.01% | 0.017% | 0.03% | 0.05% |
| Sintering Temperature | 1273K | 1273K | 1273K | 1273K | 1273K | 1273K |

| Samples | $x = 0.2$ | $x = 0.25$ | $x = 0.50$ | $x = 0.75$ | $x = 1.0$ |
|---|---|---|---|---|---|
| Composition | $La_4Ni_{2.80}Al_{0.20}O_{10}$ | $La_4Ni_{2.75}Al_{0.25}O_{10}$ | $La_4Ni_{2.5}Al_{0.5}O_{10}$ | $La_4Ni_{2.25}Al_{0.75}O_{10}$ | $La_4Ni_2Al_1O_{10}$ |
| Content of Aluminum (%Al) | 0.067% | 0.08% | 0.167% | 0.25% | 0.33% |
| Sintering Temperature | 1273K | 1273K | 1273K | 1373K | 1473K |

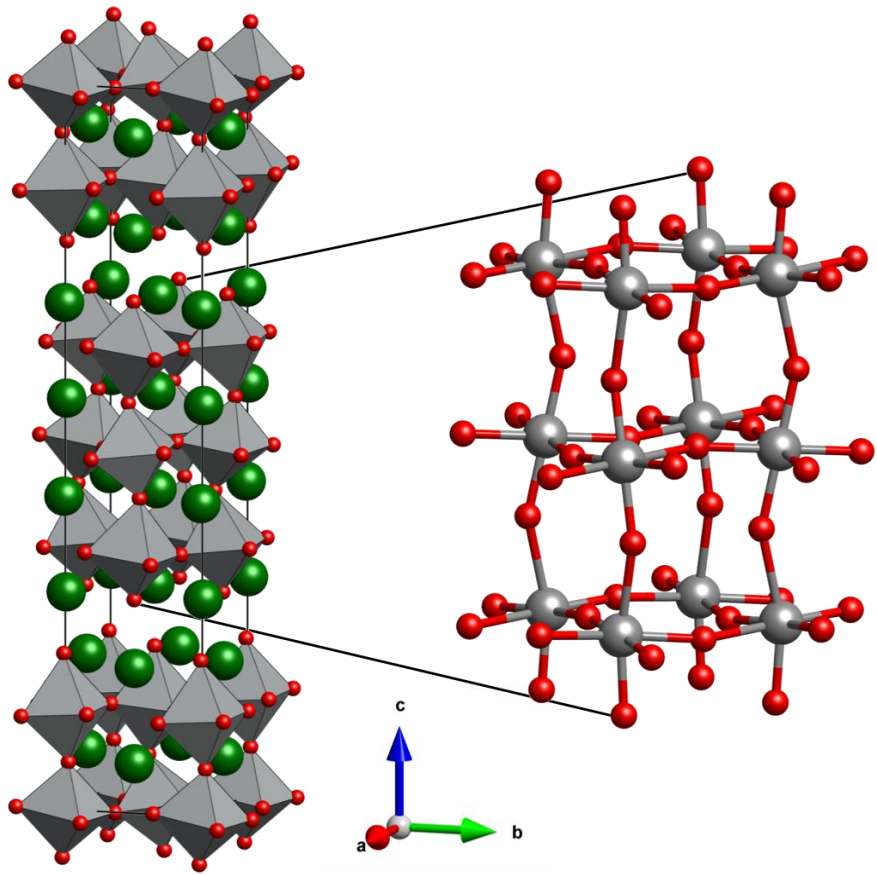

**FIG. S1.** Crystal structure for the Ruddlesden-Popper RP3 compound La$_4$Ni$_3$O$_{10}$ in space group $P2_1/a$. Green, gray and red atoms corresponds to lanthanum, nickel and oxygen respectively.

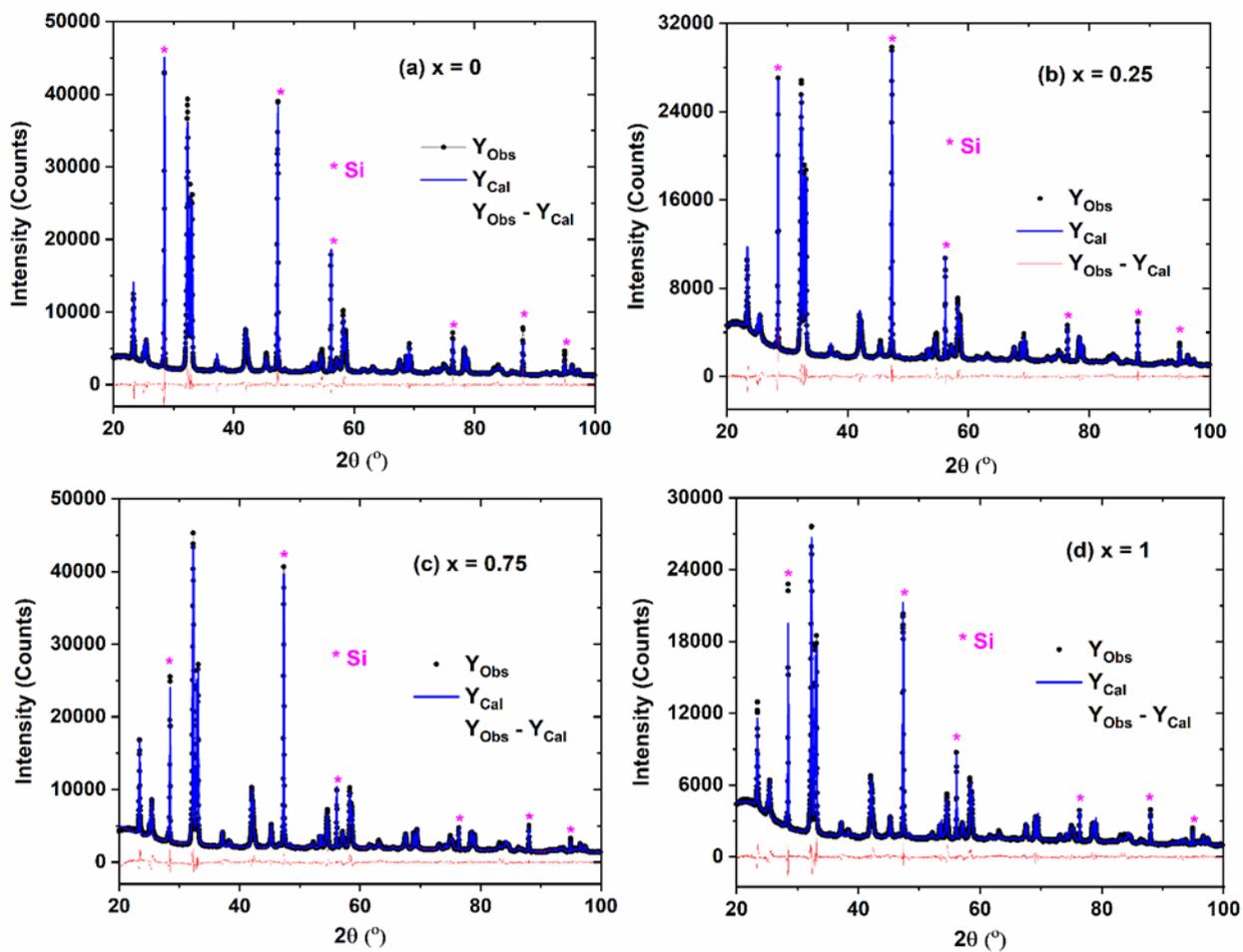

**FIG. S2.** Rietveld fit to home lab powder X-ray diffraction data ($\lambda = 1.5406$ Å) of the RP3 phases La$_4$Ni$_{3-x}$Al$_x$O$_{10}$ ($x = 0.00 – 1.00$; approximated in space group *Bmab*).

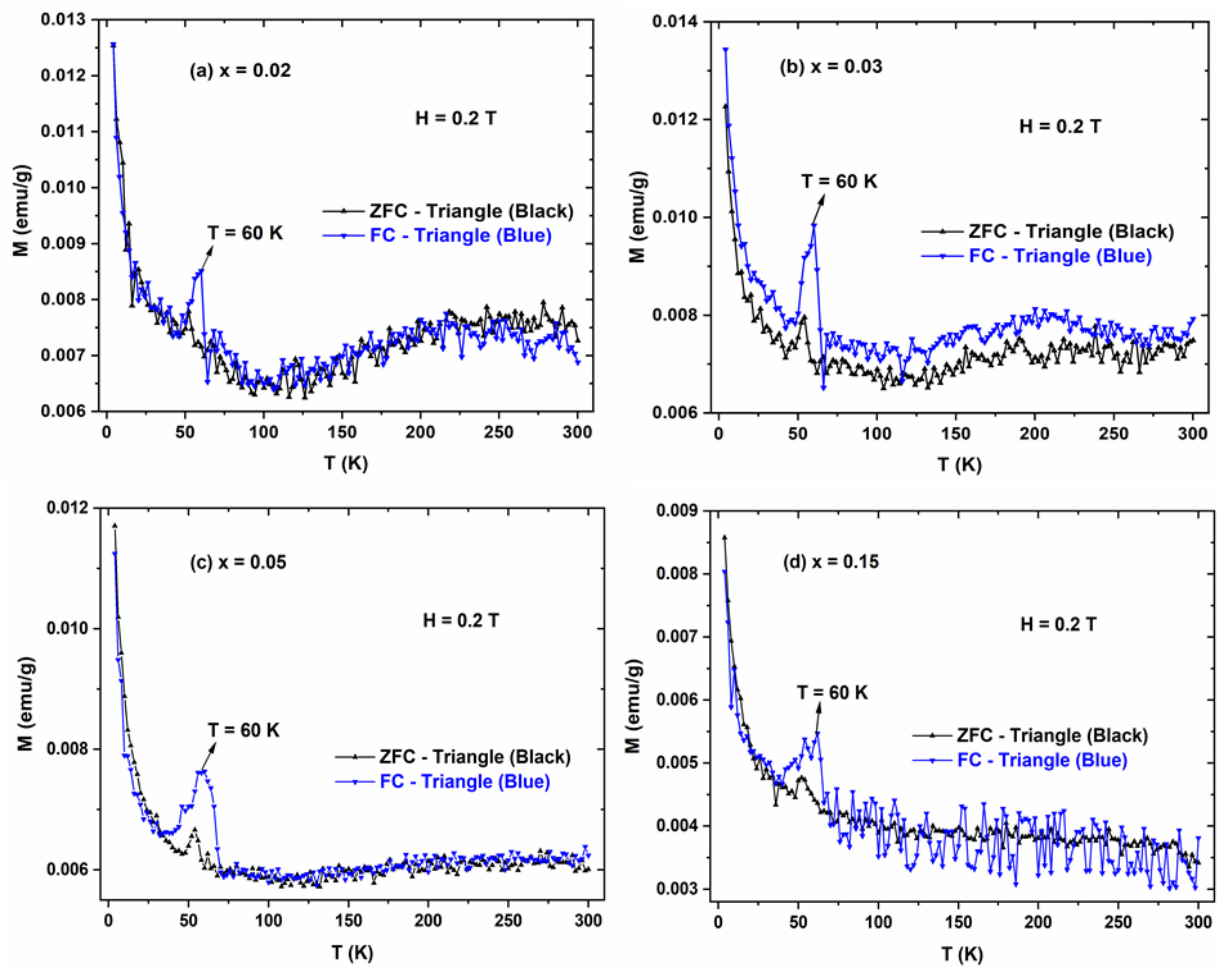

**FIG. S3.** M(T) magnetization data for $La_4Ni_{3-x}Al_xO_{10}$. Compositions ($x$) are given on the individual graphs. Anomalies at some 60 K are due to oxygen, see main text.